\documentclass[floatfix,tightenlines,showpacs,aps,twocolumn]{revtex4}
\usepackage[dvips]{graphicx} \usepackage{bm} 

\begin{document}

\title{Scanning tunnelling spectroscopy of electron resonators}

\author{J{\"o}rg Kliewer} \altaffiliation{Present address: Infineon
Technologies AG, Postfach 800949, D--81609 M{\"u}nchen}

\author{Richard Berndt} \email{berndt@physik.uni-kiel.de}
\affiliation{Institut f{\"u}r Experimentelle und Angewandte Physik\\
Christian-Albrechts-Universit\"at zu Kiel, D-24098 Kiel, Germany}

\author{S. Crampin} \email{s.crampin@bath.ac.uk} \affiliation{Department of
Physics\\ University of Bath, Bath BA2 7AY, United Kingdom\\}

\date{\today}

\begin{abstract}
The electronic structure of artificial Mn atom arrays on Ag(111) is
characterized in detail with scanning tunnelling spectroscopy and
spectroscopic imaging at low temperature. We demonstrate the degree to
which variations in geometry may be used to control spatial and spectral
distributions of surface state electrons confined within the arrays, how
these are influenced by atoms placed within the structure and how the
ability to induce spectral features at specific energies may be exploited
through lineshape analyses to deduce quasiparticle lifetimes near the Fermi
level. Through extensive comparison of $dI/dV$ maps and spectra we
demonstrate the utility of a model based upon two-dimensional $s$-wave
scatterers for describing and predicting the characteristics of specific
resonators.

\end{abstract}


\maketitle

\section{Introduction}

A central aim of nanoscale science is to controllably modify material
properties on a nanometre scale.  One of its greatest achievements has been
manipulation of surface atomic structure with the scanning tunnelling
microscope (STM) \cite{Eig_Nat90}, and some of the most striking
demonstrations of this capability the construction of complex arrays of
adatoms by the maneuvering and positioning of individual adatoms
\cite{Cro_Sci93}. Synthesized nanoscale structures such as these, as well
as other surface nanostructures including islands
\cite{Avo_Sci94,Li_PRL98_2} and step arrays \cite{Bur_PRL98} act as
resonators for mobile surface electrons, with characteristics that may be
controlled by the adjustment of geometrical parameters.

Here we present a detailed analysis of the electronic structure of a
particular class of resonator, surface nanocavities constructed from Mn
atoms on Ag(111). Of particular interest is the extent to which a simple
scattering model, which takes into account the atomic structure of the
resonator, can successfully describe its spatial and spectral electron
distributions. Recent work \cite{agmn:prl} has shown that the specific
characteristics of individual resonators impact upon the electronic
structure of adsorbates positioned within them, and may open up the
possibility of atomic scale control of magnetic or chemical properties.

In addition, we consider the use of these artificial nanoscale structures
for probing quasiparticle lifetimes close to the Fermi energy. The
lifetimes of excitations enter the description of many important surface
phenomena such as the dynamics of charge and energy transfer, and for
excitations in surface bands reflect a complex combination of decay
processes via both surface-localised and extended states, with the
additional influence of screening that varies  rapidly as the electron
density decreases outside the surface. Successful prediction of lifetimes
remains an on-going theoretical problem \cite{ECHE}, whilst measurements
must overcome considerable challenges (reviewed in Ref.\
\cite{Matzdorf_SSR98}) so that only recently \cite{Kliewer_Science00} has a
consistent account of the lifetime of a prototype quasiparticle, a hole in
a surface band, been constructed, based upon tunnelling spectroscopy at the
surface state band edge on noble metal (111) faces \cite{Li_life}. The
spatial decay of electron interference patterns observed in STM images
provides an alternative approach first indicated by Avouris~\cite{PROPO},
and has been used to measure lifetimes well removed from the Fermi energy
$E_F$ \cite{Buergi_PRL99}.  However, lifetime measurements for states close
in energy to $E_F$ are still lacking but are of particular interest because
the relative contributions of phononic and electronic decay channels are
expected to change significantly in this range. We indicate how electron
resonators may be used for this purpose.

\section{Methods}

\subsection{Experimental Procedures}

We study nanocavities constructed using Mn adatoms on the Ag (111) surface
in a custom-built ultra-high vacuum (UHV) STM, operating at a temperature
of $T=4.6$~K \cite{kliewerdiss}. The Mn atoms were evaporated onto a cold
Ag substrate that had been prepared by standard procedures in UHV.
Electrochemically etched W tips were prepared in UHV by sputtering and
annealing. Differential conductivity ($dI/dV$) spectra were recorded
under open feedback loop conditions using a lock-in amplifier with a
sinusoidal modulation added to the sample bias. Moreover, maps of $dI/dV$
were recorded simultaneously with constant current imaging without opening
the feedback.

\begin{figure}[!t] \begin{center}
\includegraphics[angle=0,width=\linewidth]{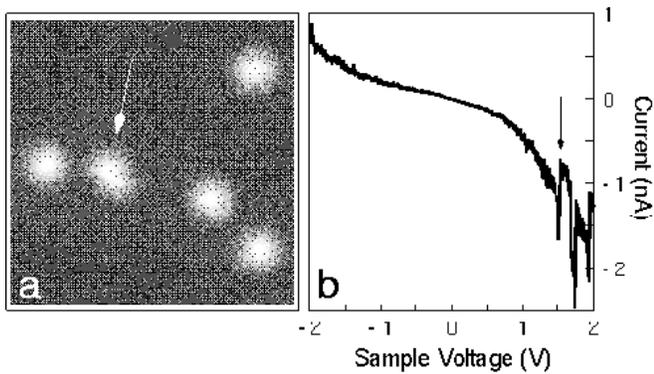} \end{center}
\vspace{-.5cm} \caption{(a) 60~$\times$~60~{\AA}$^2$ constant current
topograph of Mn adatoms on Ag(111).  At $V=$~512~mV and $R=$~731~M$\Omega$
the Mn atoms occasionally exhibit tip-induced mobility (arrow).  Image
scanned at 1.9~s/line.  (b) $I/V$ spectrum acquired on top of a single Mn
atom. Abrupt current fluctuations occur at $V=$~1.5~V and $R=$~1~G$\Omega$.
Spectrum recorded at 133~mV/sec, starting at $V=-2$~V.}
\label{atom_stability} \end{figure} 

Manganese atoms on Ag(111) image
as $1.6$~{\AA}~high and $11$~{\AA}~wide protrusions (full width at half
maximum) at a sample voltage $V=1$~V (Fig.\ \ref{atom_stability}a), similar
to previous observations of adatom species on noble metal
surfaces~\cite{Cro_Sci93,Cro_PRB93}. Most of the STM images presented here
have been recorded at $V=1$~V and a tunnelling current $I=0.01$~nA, i.e.\ a
tunnelling resistance $R=100$~G$\Omega$; these values are chosen because
the tip-sample interaction is sufficiently strong for $R$ below
$700-750$~M$\Omega$ that atoms are occasionally moved during scans, as
demonstrated in Figure~\ref{atom_stability}a. In $I(V)$ spectroscopy of
single Mn atoms we observe abrupt current changes at $V \approx 1.5$~V and
$R \approx 1$~G$\Omega$, Figure~\ref{atom_stability}b, indicating the onset
of induced atomic mobility.
To achieve acceptable signal-to-noise ratios in
tunnelling spectroscopy elevated modulation amplitudes were required
($5$~mV$_{\rm pp}$).

We construct artificial structures via controlled relocation of individual
Mn atoms, using a sliding process pioneered by Eigler \cite{Eig_Nat90}. A
Mn adatom is moved by placing the tip over the atom at large tip-sample
separation (typically $R=100$~G$\Omega$), reducing this separation
($V=3$~V, $R=500$~M$\Omega$) and then moving the tip slowly across the
surface (a rate of $\sim$ one lattice constant per second) to the desired
location. The adatom follows the tip motion, and is left in position by
retracting the tip.

\subsection{Modelling}

To model the experiments it is important to take into account the
constant-current mode of operation that is used in obtaining the $dI/dV$
maps. The tunnelling current is calculated as \cite{Selloni_PRB85}
\begin{equation} I(V,{\bm r}) \propto \int \limits_{0}^{eV} \rho_s({\bm
r},E) T(E,eV,z) dE, \label{Tunnelcurrent} \end{equation} where $\rho_s({\bm
r},E)$ is the local density of states (LDOS) of the sample at the position
${\bm r}=(x,y)$ of the tip and where the transmission probability $T$ is
given by \begin{equation} T(E,eV,z) = \exp\left[ -z
\sqrt{\frac{4m}{\hbar^2} \left( \phi_t + \phi_s + eV -2E \right)} \quad
\right] \label{Transmissionfactor} \end{equation} with tip-sample distance
$z$ and workfunctions $\phi_t=4.55$~eV and $\phi_s=4.74$~eV of tip and
sample respectively \cite{Mic_JAP77}. Simulating the feedback loop, $z$ was
adjusted for each lateral position ($z=z({\bm r})$) to ensure constant
current $I$, which in turn was numerically differentiated to obtain
$dI/dV$. In using (\ref{Tunnelcurrent}) we assume an electronically
featureless tip, which is supported by the absence of any tip dependence to
be seen during the experiments.

For $\rho_s$ we consider an array of $N$ two-dimensional $s$-wave
scatterers with coordinates $\bm R_j$, $j=0,1,\dots N-1$. This is the model
introduced by Heller {\it et al} \cite{Hel_Nat94}.  We determine $\rho_s$
from the Green function \begin{equation} \rho_s({\bm r},E)=-(1/\pi){\rm
Im}~G({\bm r},{\bm r};E) \end{equation} which is given by
multiple-scattering theory in terms of the free-electron Green function
\cite{economou} \begin{equation} G_0({\bm r},{\bm r}';E)=-{im^\ast\over 2
\hbar^2} H_0^{(1)}(\kappa|{\bm r}-{\bm r}'|) \label{FreeG} \end{equation}
where $\kappa=\sqrt{2m^\ast (E-E_0)/\hbar^2}$ and the $t$-matrices that
characterise the point scatterers: \begin{eqnarray} G({\bm r},{\bm
r}';E)&=&G_0({\bm r},{\bm r}';E) \nonumber \\ &&{}+\sum_{j,k} G_0({\bm
r},{\bm R}_j;E)T^{jk}G_0({\bm R}_k,{\bm r}';E)~~~ \label{G}\\ T^{jk}(E)&=&t
\delta_{jk}+t\sum_{l\ne j}G_0({\bm R}_j,{\bm R}_l;E)T^{lk}(E). \label{T}
\end{eqnarray}

\begin{figure*}[t] \begin{center}
\includegraphics[angle=0,width=\linewidth]{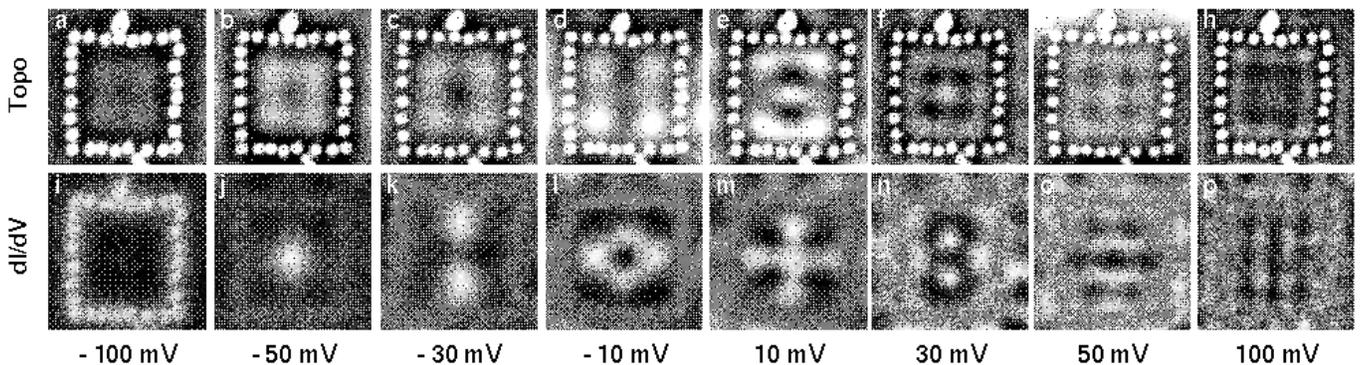} \end{center}
\vspace{-.5cm} \caption{Array of $28$ Mn atoms forming a rectangle of size
$ \approx 90 \times 100$~{\AA}$^2$. Panels (a)--(h) show constant current
topographs recorded at the indicated sample voltages. Panels (i)--(p) are
$dI/dV$ maps recorded simultaneously.} \label{box_pics} \end{figure*} 

A number of parameters enter the calculations. We use $t$-matrices
corresponding to perfectly absorbing scatterers, as identified by Heller
{\it et al}. \cite{Hel_Nat94} and used by them to describe the $dI/dV$
spectrum of a 60 Fe atom corral on Cu(111). This boundary condition
reflects the strong coupling to bulk states \cite{Cra_PRL94} that occurs at
the adatoms.  The effective mass $m^\ast = 0.42$~$m_e$ and surface state
band edge $E_0 = -67$~meV have been determined
previously~\cite{Li_PRB97,Kliewer_Science00}.  The positions of the atoms
and the lateral position of the tip are crucial in determining the exact
shape of $dI/dV$ spectra. For example, only the highest symmetry
eigenstates contribute at the centre of a circular structure, but slightly
off centre or if the shape is not exactly regular a mixing of state occurs,
giving additional structure. Uncertainty enters these coordinates through
the piezo calibration and possible anisotropy of the piezo sensitivity in
the fast ($x$) and slow ($y$) scanning directions. Therefore an
optimisation scheme was used based upon a $\chi^2$ fit using the
Levenberg-Marquardt method in which the relative coordinates were fixed but
independently scaled in the $x$ and $y$ directions. After the initial
determination, the piezo calibration parameters were held fixed in all
subsequent calculations. In a second stage, the calculations have been
extended further to include an energy dependent lifetime of the electronic
states (Sect.\ \ref{qwertz}). This was accomplished via a parameterised
imaginary self energy $i\Sigma(E)$ added to the energy $E$ in computing $G$
from equations (\ref{FreeG})-(\ref{T}). The importance of the self energy
in determining the peak amplitudes and widths in the LDOS has been noted
previously \cite{Cra_PRL94,Cra_PRB96}. Here we exploit this dependence to
extract $\Sigma$ from experimental data.

\section{Results}

\subsection{28 atom rectangle} \label{28atom_box}

Figure~\ref{box_pics} displays data from an array of $28$ Mn adatoms.
Images a--h are constant current topographs covering a range of different
voltages, recorded at values indicated in the figure. The adatoms, which
appear as bright protrusions in every case, have been arranged into a
rectangular geometry with approximate dimensions of $90 \times
100$~{\AA}$^2$. Inside this structure a weak corrugation is found,
originating from the confinement of the surface state electrons.  A
nonlinear height scale has been used for display, in order to render this
weak corrugation easily discernible; this has the side effect that the
adatoms appear rather blunt. At negative voltages (i.e. when tunnelling
from occupied levels in the sample) the constant current topographs exhibit
a rectangular pattern of four maxima within the confining array of Mn
atoms, which is still discernible at voltages well below the bottom of the
surface state band at $E_0=-67$~meV. This pattern persists owing to the
preferential tunnelling from occupied states at the Fermi level. At
positive voltages, imaging unoccupied states, the pattern becomes more
complex and varies with $V$, the most significant tunnelling now occuring
at energies corresponding to the varying bias voltage.

Images i--p show $dI/dV$ maps that were recorded simultaneously with the
topographic scans.  Below $E_0$, the adatoms cause clear features, whereas
no corrugation is discernible inside the array. In contrast, above $E_0$ a
clear wave pattern arises within the structure, while the adatoms are only
weak features which conveniently serve to indicate the rectangular
boundary. The evolution with energy of the internal pattern is as expected,
representing a progression through the nodal structures of the lowest lying
``particle-in-a-box'' eigenstates of the rectangular cavity,
$|\psi_{n,m}(x,y)|^2\propto |\sin(n\pi x/X)\sin(m\pi y/Y)|^2$ where $X$ and
$Y$ are the linear dimensions of the structure. At low energy, Fig.\
\ref{box_pics}j, a single maximum occurs ($n,m=1,1$). Next, two maxima
appear along the longer axis of the rectangle (Fig.\ \ref{box_pics}k),
corresponding to the $n,m=1,2$ level. The $-10$~mV map (Fig.\
\ref{box_pics}l) showing four maxima, two along each axis of the rectangle,
reflects contributions from both the second ($n,m=1,2$) and third
($n,m=2,1$) levels. The states are in reality resonances that in general
overlap in energy, so that $dI/dV$ maps can contain contributions from
several. At higher energies, the pattern becomes increasingly complex.

\begin{figure*}[!htb] \begin{center}
\includegraphics[angle=0,width=\linewidth]{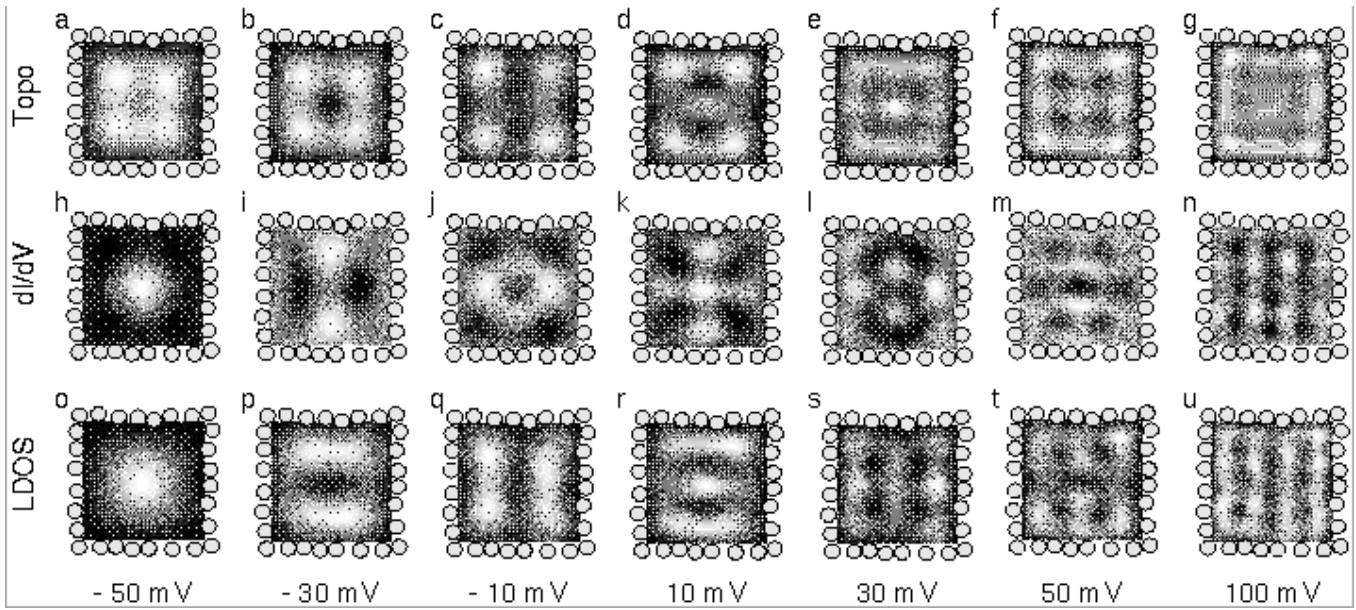} \end{center}
\vspace{-.5cm} \caption{Calculated results for the 28 atom rectangle.
(a)--(g) topographs, (h)--(n) $dI/dV$ maps, (o)--(u) LDOS.}
\label{box_theo} \end{figure*} 
Calculated results are displayed in
Fig.\ \ref{box_theo} where panels a--g are topographs, h--n are $dI/dV$
maps and o-u are LDOS maps. Close inspection and comparison with Figs.\
\ref{box_pics}b-h, and \ref{box_pics}j-p reveals that the experimental
features are reproduced well by the model calculations.  As pointed out
previously~\cite{Li_PRB97}, experimental $dI/dV$ data recorded by the
methods used here are affected by the variations in tip-sample separation
which give rise to the corrugation in the topographs.  Consequently $dI/dV$
maps are not directly comparable to LDOS maps, a fact clearly visible by
comparing Figs. \ref{box_theo}l and s, or \ref{box_theo}m and t, although
the similarity between the two tends to increase with increasing bias
voltages.  Indeed, around zero bias it is topographs and not $dI/dV$ that
are are expected to resemble the LDOS, and this is clearly the case in Fig.
\ref{box_theo}.

\begin{figure}[b] \begin{center}
\includegraphics[angle=0,width=\linewidth]{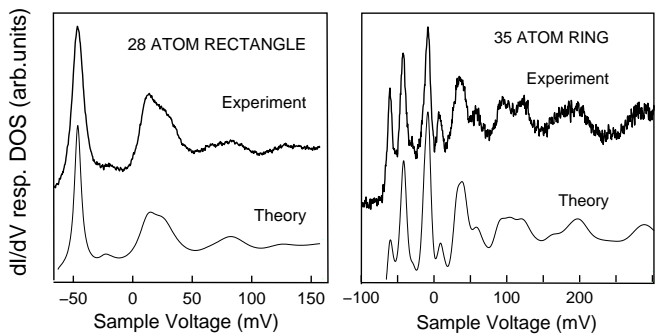} \end{center}
\vspace{-.5cm} \caption{Left: Measured $dI/dV$ spectrum and calculated LDOS
of the 28~atom rectangle. Right: $dI/dV$ spectrum and theoretical LDOS of a
$220$~{\AA}~diameter ring of 35 Mn adatoms.} \label{box_spex} \end{figure}
A more detailed comparison between experiment and theory is possible in the
case of $dI/dV$ spectra. An example measured near the centre of the
rectangular resonator is shown in the left panel of Fig.\ \ref{box_spex}.
Tunneling from a $s$-state tip directly at the centre of an ideal
rectangular structure can only take place into surface state levels with
odd quantum numbers $n,m$. This is the origin of the large peaks at $-50$
meV and between $0$ and $40$~meV. Irregularities in the structure and
deviations in position of the tip from the the exact centre introduce
additional weak tunneling channels, e.g. near $-25$ meV, whilst the
intrinsic width of the confined levels further modifies the spectrum.
Comparing the measured spectrum with the LDOS calculated using the
two-dimensional scattering model that includes the structural
irregularities and correct tip position, displayed in the same panel in
Fig.\ \ref{box_spex}, we see the latter faithfully reproduces the
differential conductivity spectrum. Note that the characteristic
temperature associated with Kondo behaviour in Mn/Ag alloys is $T_K \simeq
40$~mK~\cite{Flo_PRL70} and thus does not affect the line shape
\cite{kondo} under the present experimental conditions.

\subsection{35~atom ring}

We have also performed similar experiments and calculations on other
geometries, such as adatom rings. The right panel of Figure~\ref{box_spex}
contains a comparison between the differential conductivity spectrum,
measured approximately $5$~{\AA}~off centre in a $220$~{\AA}~diameter ring
formed from 35 Mn adatoms, and the corresponding LDOS obtained using the
$s$-wave scattering model. The spectrum contains a large number of peaks,
the off-centre position enabling low symmetry eigenstates to provide
channels for tunnelling, but these are reproduced in the theoretical LDOS
spectrum calculated using the precise atomic positions of the Mn ring. We
conclude that the $s$-wave scattering model provides an adequate basis for
modelling the spatial and spectral electron distributions in electron
resonators.

\subsection{Adatoms in resonators}

\label{Sec_spectro_full}

\begin{figure}[htb] \begin{center}
\includegraphics[angle=0,width=\linewidth]{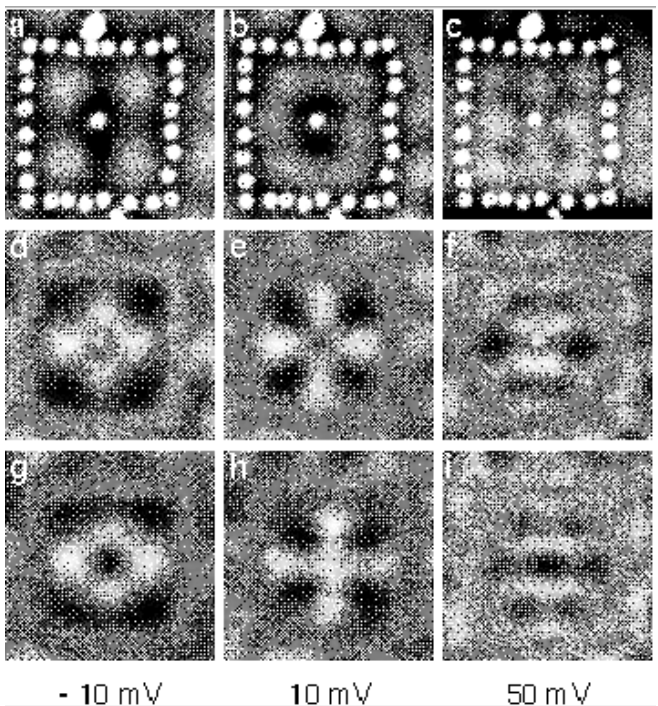} \end{center}
\vspace{-.5cm} \caption{Topographs of a 28 atom rectangle with a central Mn
adatom recorded at the indicated sample voltages (a-c, top row), and
$dI/dV$ maps of the same structure with (d-f, middle row) and without (g-i,
bottom row) the central atom.} \label{full_empty} \end{figure}

Having
characterised the properties of electron resonators we have next
investigated the interaction with an additional atom introduced into the
nanocavity region. These structures were formed by maneuvering one of the
boundary Mn atoms into the central region, and then moving a Mn atom
elsewhere on the surface into the vacant position in the boundary,
restoring the integrity of the resonator.

In Figure \ref{full_empty} we show $dI/dV$ maps of a rectangular structure
made from 28 Mn atoms and measured at three different voltages, contrasting
the images obtained with and without a Mn atom inside the structure. At
first glance the presence of the central atom appears to barely affect the
wave pattern inside the rectangle, apart from in the immediate vicinity of
the centre atom. There, the difference is striking.  At voltages of
$V=-10$~mV and $V=50$~mV where $dI/dV$ is at a minimum in the empty
structure (Figs.\ \ref{full_empty}g and i), $dI/dV$ exhibits a peak in the
centre upon insertion of the atom (Figs.\ \ref{full_empty}d and f). At
$V=10$~mV, however,  the maximum observed in the empty rectangle (Fig.\
\ref{full_empty}h) changes to a minimum (Fig.\ \ref{full_empty}e).

\begin{figure}[!ht] \begin{center}
\includegraphics[angle=0,width=\linewidth]{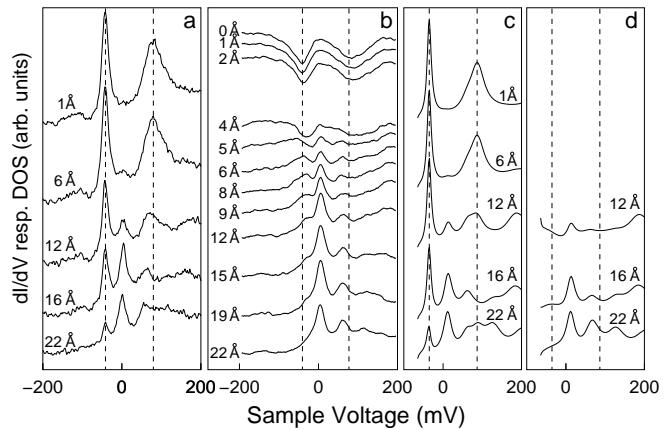} \end{center}
\vspace{-.5cm} \caption{Evolution of recorded $dI/dV$ spectra with
increasing lateral distance from the centre of a 16~Mn atom ring radius
$\sim 40$~\AA~ (a) without and (b) with a central Mn atom, and spectra
calculated using the two-dimensional scattering model of the LDOS for the
same systems, (c) without and (d) with the central atom. The vertical
dashed lines mark the energies of the main spectral features present at the
centre of the empty structure.} \label{abstands_kurve80} \end{figure}

This ``contrast reversal'' has been previously reported and explained as a
consequence of the strong hybridisation between states on the Mn atom and
substrate surface electron levels~\cite{agmn:prl}, and has been reproduced
in calculations of the local density of states directly on Mn atoms in
resonators of various geometries using full three-dimensional multiple
scattering calculations. Here we consider how this interaction extends away
from the central Mn atom and the degree to which the two-dimensional
scattering model describes the behaviour. Fig.\ \ref{abstands_kurve80}
shows a series of $dI/dV$ spectra measured at various distances from the
centre of an empty adatom ring, and when a Mn atom has been introduced.
Also shown are the LDOS calculations for identical geometries. Comparing
$dI/dV$ spectra in Figs. \ref{abstands_kurve80}a and b at central positions
closest to the location of the new adatom, it is clear that spectral peaks
in the empty resonator coincide with troughs of the occupied structure
(vertical dashed lines).  This is evidence for strong hybridisation
\cite{agmn:prl}. Moving laterally away from this central position the
characteristic peak structure of the empty resonator returns, with the
exception of the lowest peak near $-40$~mV which remains absent even at
large displacements.  This prominent feature is associated with a nodeless
resonator eigenstate that has its maximum at the centre of the ring, and so
is particularly sensitive to the introduction of a Mn atom into the central
region.

The two right-most panels in Fig.\ \ref{abstands_kurve80}  show the LDOS
calculated for this system using the two-dimensional scattering model, at
lateral positions at which $dI/dV$ spectra were recorded. The prominent
features in the experimental spectra are reproduced well in the theoretical
curves, again showing that the simple scattering model, using the exact
atomic locations of the Mn atoms making up the structure, gives an accurate
and valid account of the electronic structure of resonators. The
calculations also confirm that the dominant state in the empty resonator is
absent in the spectrum of states of the occupied resonator, so that the
insensitivity of $dI/dV$ maps to the presence of the central atom away from
its immediate vicinity that was suggested by Fig. \ref{full_empty} clearly
does not extend to all sample voltages. Note that in Fig.
\ref{abstands_kurve80}d we have omitted theoretical spectra within
$10$~\AA~ of the central Mn atom. Within this radius (which is much greater
than the topographic radius of the Mn atoms) the differential conductivity
also has large contributions from channels involving bulk states scattered
by the adatom \cite{buried}, and these are not included in the model. Also
omitted in the model is any spatial information about the adatom orbitals
that would also be required to model $dI/dV$ spectra (and maps) in its
immediate vicinity.

\subsection{Energy dependent self energy}

\label{qwertz}

The basic idea of our approach to lifetime measurements is straightforward
\cite{kliewerdiss}.  Exploiting the properties of the electron resonators
identified in the previous sections we construct nanocavities with
electronic features in the vicinity of $E_F$. The lineshape of the spectral
features are affected by lifetime-limiting processes so that through
analysis the lifetime $\tau$ can be determined. By changing the geometry of
nanoscale structures it is possible to tune the eigenstate energies of the
structures, and thus generate states at various energies in order to obtain
the energy dependence of the lifetime. The two-dimensional scattering model
validated above enables the design of suitable structures. We note that a
fully quantitative analysis will also require information on the energy
dependence of the tip density of states, which is not expected to be
constant over a wide energy range, and likewise the bulk density of states
of the crystal which causes a background signal in $dI/dV$ spectra. Here,
we seek to demonstrate the concept and so neglect these influences.
Following Li et al. \cite{Li_life} we describe the lifetime effects in
terms of an effective imaginary part $\Sigma$ of the self energy, where
$\tau = \hbar/(2 \Sigma)$.

\begin{figure}[t] \begin{center}
\includegraphics[angle=0,width=\linewidth]{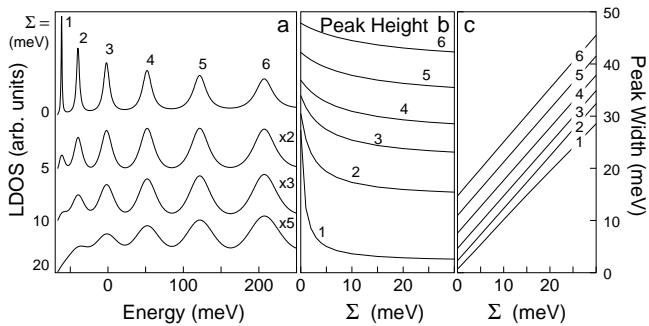} \end{center}
\vspace{-.5cm} \caption{Effect of a constant imaginary part $\Sigma$ of the
self energy on the LDOS of a 36 atom circle with 210~{\AA}~ diameter. (a)
Calculated LDOS at the centre for various $\Sigma$ indicated in the figure.
Variation of peak (b) heights and (c) widths with $\Sigma$.}
\label{ei_theory} \end{figure}

The effect of $\Sigma$ on the LDOS of
an artificial nanoscale resonator is shown in Fig.\ \ref{ei_theory}a for a
210~{\AA}~ diameter ring.  The calculation for vanishing $\Sigma$, i.e.\
infinite lifetime, yields a series of sharp resonances (Fig.\
\ref{ei_theory}a, top curve).  Note that the spectrum corresponds to the
centre of the structure. The high symmetry gives the ``cleanest'' spectrum,
avoiding difficulties that exist at off-centre positions where resonances
can overlap. With increasing $\Sigma$ there is a reduction of the peak
heights and a concomitant broadening, especially apparent for the lowest of
the states which have the sharpest intrinsic profile. These trends are
quantified by fitting a series of Lorentzians to the data, and
Figure~\ref{ei_theory}b and c display the widths and height extracted
through this analysis. As $\Sigma$ increases, the peak heights decrease
monotonically, the lowest (in energy) peaks most rapidly, whilst there is
an essentially linear dependence of the peak width on $\Sigma$.

\begin{figure}[t] \begin{center}
\includegraphics[angle=0,width=0.9\linewidth]{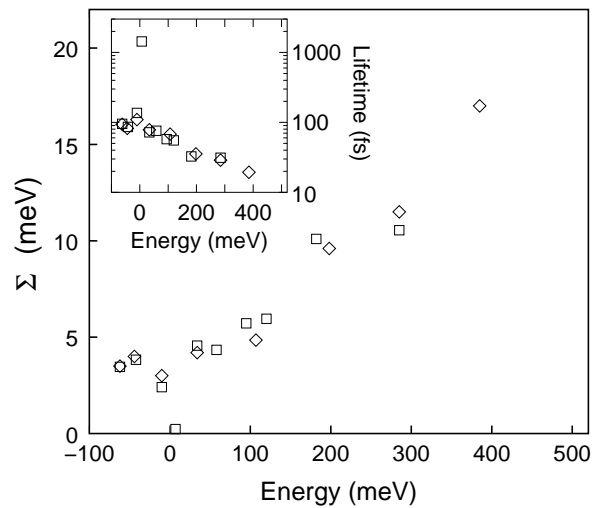} \end{center}
\vspace{-.5cm} \caption{$\Sigma(E)$ extracted from analysis of $dI/dV$
spectra of a 115~\AA~ radius 35~atom circular resonator (diamonds, squares)
and a 40~\AA~ radius 16~atom circular resonator.
The $\Sigma$ values are plotted at the position of the
respective spectral features. The inset shows the equivalent lifetimes
$\tau=\hbar/(2 \Sigma)$ of hot holes and electrons.} \label{ei_final}
\end{figure}

Figure~\ref{ei_final} shows values of $\Sigma(E)$
obtained from an analysis of measurements from two nanostructures. For
each, $\chi^2$ fits were made over limited energy ranges ($\sim 100$ meV)
adjusting coefficients of a linear parameterisation of $\Sigma$. We
estimate the error of $\Sigma$ thus determined to be of the order of $\pm
30$ \%. The diamonds and squares are derived from $dI/dV$ spectra for a
35~atom circular resonator, and the filled circles were obtained from a
16~atom circle.  Below the Fermi energy ($E=0$) $\Sigma$ approaches a value
obtained previously from spectra taken on clean terraces which yielded
$\Sigma=3$ meV \cite{Kliewer_Science00}. There is a clear increase of
$\Sigma$ at large energies, and we find $\Sigma(E=0.5$~eV$) \sim 20$~meV,
which is comparable with an extrapolation down to this energy of data for
the Ag(111) surface state extracted from the spatial coherence of
interference patterns at energies $E>1$~eV by B\"urgi {\it et al.}
\cite{Buergi_PRL99}. Finally, at energies close to $E_F$ we observe a
marked decrease of $\Sigma$ which corresponds to an increased lifetime of
the corresponding quasiparticles (inset in Fig.\ \ref{ei_final}). Increased
lifetimes are indeed expected as both the phononic and electronic decay
channels become less efficient.  While further data close to $E_F$ is
necessary to quantitatively characterize this effect the first results are
encouraging.  Future work will also address the relation of the
self-energies determined from atomic arrays to that of an unperturbed
surface.  We note that data from resonators of different diametres (40 \AA\
and 115 \AA) yield similar $\Sigma(E)$.  This similarity appears to
indicate that the influence of the array may be rather small.

\section{Summary}

Using scanning tunnelling spectroscopy and spectroscopic imaging at low
temperature we have investigated artificial Mn atom arrays on Ag(111).
These act as electron resonators and we have demonstrated the degree to
which variations in geometry may be used to control spatial and spectral
distributions of confined surface state electrons. This control may be
exploited in manipulating the local density of states of adsorbates
introduced into the resonator, or as we have demonstrated, to generate
spectral features at specific energies in order to perform lineshape
analyses from which quasiparticle lifetimes may be deduced. Extensive
comparison of $dI/dV$ maps and spectra indicated the utility of a model
based upon two-dimensional $s$-wave scatterers for describing and
predicting the characteristics of specific resonators.

\section*{Acknowledgements}

This work has been supported the DFG, the DAAD and the UK EPSRC.



\begin{thebibliography}{}

\bibitem{Eig_Nat90} D. M. Eigler, E. K. Schweizer, Nature {\bf 344}, 524
(1990).

\bibitem{Cro_Sci93} M. F. Crommie, C. P. Lutz, D. M. Eigler, Science {\bf
262}, 218 (1993).

\bibitem{Avo_Sci94} Ph.\ Avouris, I.-W. Lyo, Science {\bf 264}, 942 (1994).

\bibitem{Li_PRL98_2} J. Li, W.-D. Schneider, R. Berndt, S. Crampin, Phys.\
Rev.\ Lett.\ {\bf 80}, 3332 (1998).

\bibitem{Bur_PRL98} L. B\"urgi, O. Jeandupeux, A. Hirstein, H. Brune, K.
Kern, Phys.\ Rev.\ Lett.\ {\bf 81}, 5370 (1998).

\bibitem{agmn:prl} J. Kliewer, S. Crampin, R. Berndt, Phys.\ Rev.\ Lett.\
{\bf 85}, 4936 (2000).

\bibitem{ECHE} P.M. Echenique, J. Osma, V.M. Silkin, E.V. Chulkov, J.M.
Pitarke, Appl.\ Phys.\ A {\bf 71}, 503 (2000).

\bibitem{Matzdorf_SSR98} R.~Matzdorf,  Surf.\ Sci.\ Rep.\ {\bf 30}, 153
(1998).

\bibitem{Kliewer_Science00} J.~Kliewer, R.~Berndt, E.~V.~Chulkov,
V.~M.~Silkin, P.~M.~Echenique, S.~Crampin, Science {\bf 288}, 1399 (2000).

\bibitem{Li_life} J. Li, W.-D. Schneider, R. Berndt, O. R. Bryant, S.
Crampin, Phys.\ Rev.\ Lett.\ {\bf 81}, 4464 (1998).

\bibitem{PROPO} Ph.~Avouris, I.-W.~Lyo, R.~E.~Walkup, Y.~Hasegawa,
J.~Vac.~Sci.~Technol.~B {\bf 12}, 1447 (1994).

\bibitem{Buergi_PRL99} L.~B\"urgi, O.~Jeandupeux, H.~Brune, K.~Kern, Phys.\
Rev.\ Lett.\ {\bf 82}, 4516 (1999).

\bibitem{kliewerdiss} J.~Kliewer, PhD thesis, RWTH Aachen, D-52056 Aachen,
Germany (2000).

\bibitem{Cro_PRB93} M. F. Crommie, C. P. Lutz, D. M. Eigler, Phys.\ Rev.\ B
{\bf 48}, 2851 (1993).

\bibitem{Selloni_PRB85} A. Selloni, P. Carnevali, E. Tosatti, C. D. Chen,
Phys.\ Rev.\ B {\bf 31}, 2602 (1985).

\bibitem{Mic_JAP77} H. B. Michaelson, J. Appl.\ Phys.\ {\bf 48}, 4729
(1977).

\bibitem{Hel_Nat94} E. J. Heller, M. F. Crommie, C. P. Lutz, D. M. Eigler,
Nature {\bf 369}, 464 (1994).

\bibitem{economou} E.N.~Economou, {\it Green's Functions for Quantum
Physics} (Springer-Verlag, Berlin, 1983).

\bibitem{Cra_PRL94} S. Crampin, M. H. Boon, J. E. Inglesfield, Phys.\ Rev.\
Lett.\ {\bf 73}, 1015 (1994).

\bibitem{Li_PRB97} J. T. Li, W.-D. Schneider, R. Berndt, Phys.\ Rev.\ B
{\bf 56}, 7656 (1997).

\bibitem{Cra_PRB96} S. Crampin, O. R. Bryant, Phys.\ Rev.\ B {\bf 54},
R17367 (1996).

\bibitem{Flo_PRL70} J. Flouquet, Phys.\ Rev.\ Lett.\ {\bf 25}, 288 (1970).

\bibitem{kondo} J. Li, W.-D. Schneider, R. Berndt and B. Delley, Phys. Rev.
Lett. {\bf 80}, 2893 (1998).

\bibitem{buried} S. Crampin, J. Phys.: Condens. Matter {\bf 6}, L613
(1994).

\end{thebibliography}
\end{document}